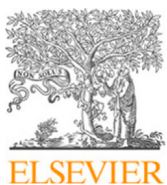
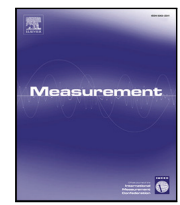

# High-energy electron measurements with thin Si detectors

Gokul Das H [a], R. Dubey [a,*], K. Czerski [a], M. Kaczmarski [a], A. Kowalska [b], N. Targosz–Ślęczka [a], M. Valat [a]

[a] *Institute of Physics, University of Szczecin, 70-451 Szczecin, Poland*
[b] *Institute of Mathematics, Physics and Chemistry, Maritime University of Szczecin, Poland*

A B S T R A C T

A technique for measuring high-energy electrons using Si detectors of various thicknesses that are much smaller than the range of the examined electrons is presented. The advantages of the method are discussed on the basis of electron–positron pair creation recently studied in deuteron–deuteron fusion reactions at very low energies. Careful Geant4 Monte Carlo simulations enabled to identify the main spectral contributions of emitted electrons and positrons resulting from the energy loss mechanisms and scattering processes within the target, detector and their holders. Significant changes in the intensity of the detected electrons, depending on the detector thickness and the thicknesses of absorption foils placed in the front of the detector could be observed. The corresponding correction factors have been calculated and can be used for different applications in basic and applied research.

## 1. Introduction

The energy deposition of electrons in matter differs strongly from that known for heavier charged particles. Due to their low mass, the ionization process of atoms by electron–electron scattering can occur at large scattering angles, leading to large path lengths compared to the extrapolated projected range [1]. The latter makes it difficult to determine a proper detector thickness for the detection of electrons of a known kinetic energy [2]. Furthermore, the intensity of the full energy line, obtained for a mono-energetic electron source measured in detectors of different thicknesses [1,3], undergoes significant changes, and the contribution of scattered electrons producing a continuous energy spectrum additionally hinders the analysis. The problems arise especially for high energy electrons for which thicker detectors are needed and the exact number of detected electrons should be determined [1]. Additionally, high-energy electrons lose their energy also by bremsstrahlung, which makes the estimations even much more complicated. The best solution is to compare the measured energy spectrum with a simulated one, assuming that all physical processes involved are taken into account [4]. Moreover, the high penetrability of high-energy electrons makes it necessary to model the entire experimental setup. For this purpose, the Monte Carlo (MC) simulation platform, Geant4 [5,6] seems to be the ideal tool.

High-energy electron measurement is crucial in various basic and applied sciences, including fundamental nuclear physics, nuclear energy research, medical imaging, radiation therapy, highly energetic light particles flux in the Earth's radiation belts [7–9]. Different measurement methods are employed based on the specific requirements of experiments and the energy range of interest (as presented in Table 1) [10–16]. Researchers often choose a combination of Si detectors with other detection techniques based on specific goals and conditions of their experiments.

In the present work, we will focus on a measurement technique that can be applied to high-energy electrons or positrons produced in very weak physical processes for which a compact measurement geometry and large detection efficiency are needed. An example for this kind of application is the deuteron fusion (DD) reactions $^2$H(d,p)$^3$H and $^2$H(d,n)$^3$He studied at very low energies, far below the Coulomb barrier. Recently, we have observed an indication of emission of high-energy electrons/positrons in the deuteron–deuteron (DD) reaction [17] from the previously predicted threshold resonance [18] in the compound nucleus $^4$He at the excitation energy of 23.84 MeV. This resonance has spin-parity, $J^\pi = 0^+$ and can predominantly decay to the $^4$He ground state by internal pair creation (IPC) giving continuous energy spectra of emitted electrons and positrons. The measurements have been performed under ultra-high vacuum conditions in a small target chamber using very compact detection geometry to increase detection probability and compensate for strongly decreasing cross-sections with the lowering deuteron energy. We have used only a single Si detector to determine the reaction branching ratio between emitted electron–positron pairs and the 3.02 MeV protons from the reaction

**Table 1**
A comparative table of measurements of high-energy electron performed with silicon (Si) detectors. The table includes information about the experimental setup, energy range, resolution, efficiency, and other relevant factors.

| Experimental setup | Energy range[a] (MeV) | Energy resolution[b] | Detector efficiency[b] | Advantages | Limitations | Ref |
|---|---|---|---|---|---|---|
| SSB/PIPS Detectors (> 300 μm Thickness) | 0.1–20 | 0.2% at 50 MeV | 99% at 50 MeV | Compact design, Wide energy- range, high efficiency | Sensitive to temperature variations | [1,3,4,10,11] |
| Si detector array (< 300 μm Thickness) | 0.01–1 | 1% at 5 MeV | 90% at 5 MeV | Compact design, high efficiency | Limited to lower energy | [11,12] |
| Pixelated Si detector | 0.001–1 | 0.1% at 500 keV | 95% at 500 keV | Excellent spatial resolution | Complex readout electronics, higher cost | [13,15] |
| Si strip detectors | 10–100 | 0.5% at 50 MeV | 80% at 50 MeV | Good energy resolution, large coverage | Saturation at high energies, temperature sensitivity | [10,14] |
| Hybrid Telescope (e.g., Si/CsI(Tl) detector) | 1–100 | 1% at 10 MeV | 92% at 10 MeV | Particles Coincidence information's | Complex hybrid setup, non-uniform response, additional noise | [10,16] |
| Si drift detector | $10^3$-$10^4$ | 0.1% at 5 GeV | 95% at 5 GeV | Excellent energy resolution, good timing | Limited to charge particles, complex design | [10] |

[a] Energy ranges are not explicit values; they may vary by 10%–20%, depending upon physics considerations.
[b] Energy resolution and detector efficiency are also not explicit values; they may vary depending on particular experimental conditions.

$^2$H(d,p)$^3$H down to the lowest possible projectile energies with high precision. Simultaneous measurements of protons and electrons using Si detectors of various thicknesses will allow us to track the differences in the detection of both charged particles.

In the other comparable studies, high-energy electrons/positrons were measured by means of large detector telescopes consisting of a thin–thick plastic scintillator [19]. Lately, a multiarray of dE-E plastic scintillator telescopes and improved dE-E hybrid (Double Sided Silicon Strip Detectors(DSSD) and plastic scintillator) telescopes were used to measure the energy and angular correlation of electron–positron pairs. These studies examined the possibility of creating a hypothetical boson with a mass of approximately 17 MeV/c$^2$ decaying via IPC in the $^3$H(p,e$^+$ e$^-$)$^4$He nuclear reaction [20]. In contrast to the studies above, we apply a relatively thin, single Si detector the thickness of which is much smaller than the range of high-energy electrons. Such detectors are typically utilized for the detection of heavy charged particles simultaneously produced in various nuclear reaction channels. The advantage of the method proposed, besides its high detection efficiency being beneficial for low cross section measurements, is also that the continuous spectrum of high energy electrons/positrons could be measured by their energy loss in the Si detector. Instead of a broad energy spectrum, we will observe only a well-defined absorption line in a small energy region, improving the effect-background ratio. Furthermore, extensive Geant4 Monte Carlo simulations will allow us to explain different contributions of the measured energy spectra resulting from scattering and energy loss processes, which has not been discussed in the literature up to now.

## 2. Experimental setup and Geant4 Monte Carlo simulation model

The experiments were performed at the Ultra High Vacuum (UHV) Accelerator Facility of the University of Szczecin, Poland [21]. The experimental setup for the DD fusion reactions, with a single silicon detector, Al foil, ZrD$_2$ target and target holder is shown in Fig. 1(a). The corresponding geometry simulated with Geant4 MC and the detailed construction of the SSB detector with Al foil simulated with Geant4 MC are displayed in Figs. 1(b) and 1(c). To ensure the measurement accuracy and reliability of the detectors and the reproducibility of simulations, the energy calibration of the detectors using a standard radioactive source (RS) was performed. Further more, to clearly separate the accelerator and reaction-induced events, comprehensive Geant4 MC simulations were performed. These simulations took into careful consideration the entirety of the experimental setup, encompassing the target, its backing material, the target holder, and all surrounding materials near the target and detectors. The simulations also accounted for the packaging of the detectors and the various components involved in mounting them. As shown in Fig. 1, the beam was impinged on a 0.5 mm thick ZrD$_2$ target plate that was tilted at 45$^0$ to the beam, resulting in a beam spot size of 7 × 12 mm. EG ORTEC surface barrier silicon detectors [22] of 0.3–3 mm thicknesses were used for detection of all charged particles emitted in the DD reactions: protons (3.02 MeV), tritons (1.01 MeV), and $^3$He particles (0.82 MeV), as well as electrons and positrons resulting from the internal pair creation (continuous energy spectrum up to 22.8 MeV). In the front of the detector, an Al absorption foil was placed to protect the detector against elastically scattered beam deuterons and to reduce energy or fully absorb emitted charged particles. The foil thickness was ranging between 5 μm to 90 μm allowing for background-free observation of electrons as explained in Section 3. The DD reactions were studied for deuteron energies below 20 keV, which means that the reaction products were mainly created at the target depth smaller than 200 nm. The detector was situated at a backward angle 135$^0$ and 7 cm away from the target.

### 2.1. Geant4 Monte Carlo simulation

The C++ based Geant4 MC simulation code has been used for simulating the transport of charged particles through matter and the response function of detectors — in average we have generated 10 million events. For the calibration of detectors, the detailed geometry of the detectors, their construction as well as the radioactive source geometry were incorporated in the simulation using the General Particle Source (GPS) function in Geant4 MC . "G4Radioactive Decay Physics" and "Electromagnetic Physics List Option 4" were also used for the simulations. For the DD reactions, the simulations were performed in two steps to generate the response function of fusion products in the detectors. In the first step, the low energy deuterons of energy 20 keV were assumed to interact with the ZrD$_2$ target, and the depth distribution of impacting deuterons up to 300 nm was determined. A custom tailored "QGSP BIC HP EMZ Physics" list with more accurate stopping power values and a very small step size (∼5 nm) was used for this purpose. In the second part of the simulation, the fusion products were emitted form planar layers placed at different target depths, with intensity scaled with the calculated cross sections. Even though the target layers may cause no impact on the high-energy electrons, the straggling and energy loss that might happen to the protons were accounted for through this procedure.

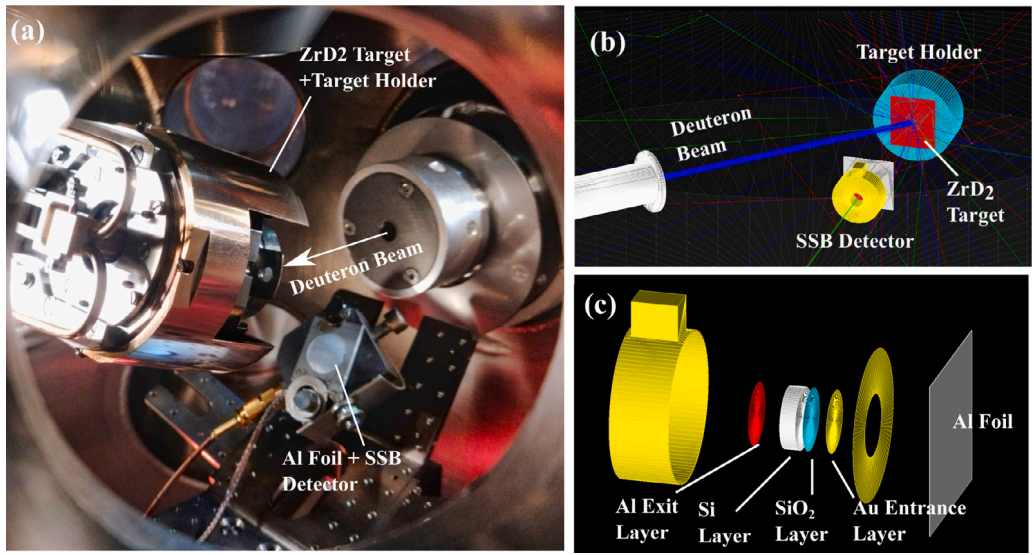

**Fig. 1.** The experimental setup for DD fusion reactions utilized in the present study, as observed through the viewport of the Ultra High Vacuum (UHV) chamber (a). The direction of the incident beam and the positions of the target, target holder, Al foil and the Silicon Surface Barrier (SSB) detector are marked. The corresponding geometry simulated with Geant4 MC, visualized with OpenGL and the included detailed construction of the SSB detector with Al foil are presented in figures (b) and (c), respectively.

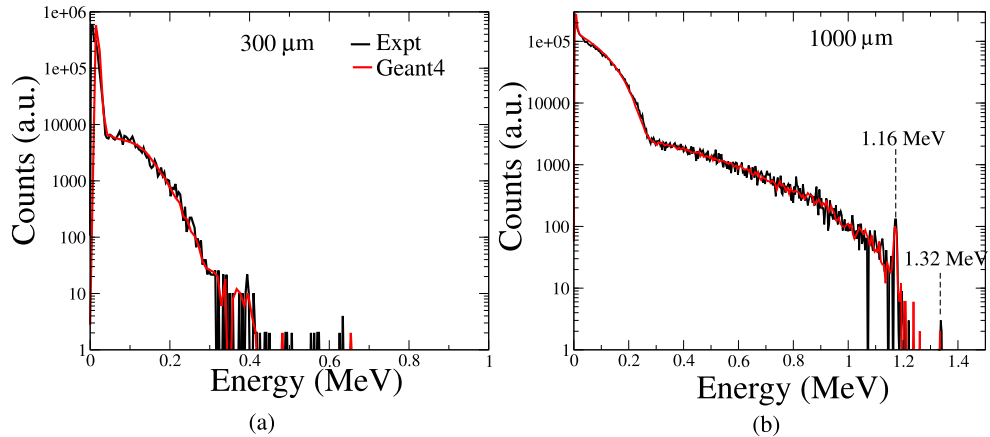

**Fig. 2.** The comparison between the experimental (black curve) and Geant4 simulated (red curve) spectra of the $^{60}$Co beta source obtained using 300 μm (a) and 1000 μm (b) thick Silicon Surface Barrier detectors. The beta endpoint energy of 317.9 keV is distinctly identifiable in both spectra. Notably, in the 1000 μm thick detector, conversion electron peaks corresponding to the 1.17 MeV and 1.33 MeV gamma rays can be observed.

## 2.2. Detector calibration with radioactive sources

The $\beta^-$ $^{60}$Co radioactive point source was positioned 4 cm from the entrance window of the detector and the measurement was carried out for a live time of 10 min. An aluminium degrader of thickness 3.65 μm was placed in front of the detectors and the measurements were repeated. Long background measurements of 240 min each were conducted before and after the measurements with the radioactive source and the averaged background spectrum was subtracted from the radioactive isotope spectra. The $^{60}$Co radioactive source exhibits two $\beta^-$ decay channels, with end-point energies of 0.317 MeV and 1.48 MeV, resulting in the decay product $^{60}$Ni. The corresponding branching ratios for these channels are 99.88% and 0.12%, respectively [23].

The beta endpoint energy of 317 keV can be prominently observed in 300 and 1000 μm thick detectors (see Fig. 2). Additionally, in the 1000 μm detector, the higher energy $\beta^-$ decay with the energy endpoint of 1.48 MeV and two distinct conversion electron peaks resulting from the 1173.23 keV and 1332.49 keV transitions are clearly visible. The differences between the measured spectra obviously arise from different energy depositions of electrons in the thinner and thicker detectors. The overall efficiency of the calibration measurements was about 95%, which results mostly from the uncertainty of the measurement geometry when compared to the known activities of the sources.

## 3. Results

Recently, it has been proposed that threshold resonance in the compound nucleus $^4$He at the excitation energy 23.8 MeV can predominantly decay to the ground state of $^4$He via internal pair creation (IPC) giving continuous energy spectra of emitted electrons and positrons [17]. To ensure a correct interpretation of the electron/positron spectra obtained in the DD reaction, it is essential to have a clear understanding of the origin of each energy spectrum component. For example, besides the IPC process, the background contributions from external pair creation (EPC) and multiple electron scattering are also possible. For this purpose, several event generators were developed and applied to the Geant4 MC code.

### 3.1. Implementation of internal pair creation in Geant4

An internal pair creation is the electromagnetic process by which a nucleus emits an $e^+e^-$ pair instead of gamma-rays, in conjunction with the recoil of the emitting nucleus. The differential IPC cross-section calculation for the $E_0$ transition in $^4$He follows the method proposed by Rose [24,25]. Using a two-dimensional random generator that is part of the HBOOK library (CERN), the emission angle and energy of a positron related to an IPC event are generated based on input to a two-dimensional random generator [24]. The energy spectrum of emitted electrons/positrons is displayed as an inset in Fig. 4(b).

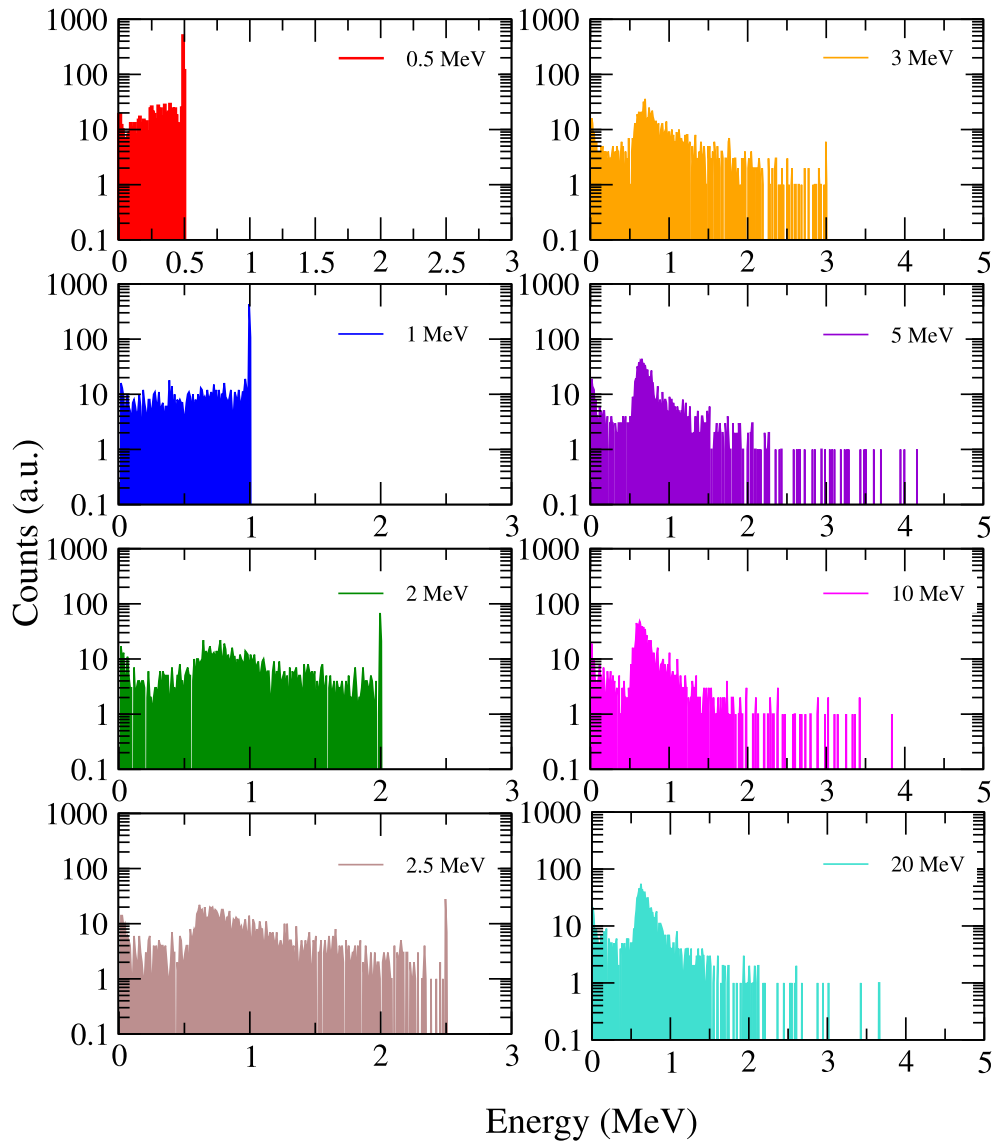

**Fig. 3.** Response function of a 2000 μm thick Si detector to monoenergetic electrons of energies ranging between 0.5 and 20 MeV, simulated using Geant4 MC. The full absorption peak diminishes with increasing electron energy. For energies greater than 3.5 MeV, the partial absorption peak is only visible. The change in the position of the partial absorption peak for lower electron energies can also be observed.

### 3.2. Energy spectrum of electrons

The response function of the Si detector calculated for monoenergetic electrons using Geant4 MC simulations (see Fig. 3) exhibits two prominent peaks, namely the full energy peak resulting from the entire absorption of $e^+e^-$ particles, and an average (partial) energy loss peak arising from the incomplete absorption of energy. When an electron with energies below 2.5 MeV traverses the 2 mm silicon (Si) detector, the observed spectra are mostly characterized by the presence of the full energy peak [3]. However, there is a tailing effect observed at lower energies due to partial energy deposition in the detector. For higher electron energies, the full energy peak gets gradually weaker and disappears for 5 MeV electrons, completely. Instead of that, a broad bump at energies below 1 MeV can be observed, the position of which does not depend on the incident electron energy. This structure corresponds to the average energy loss of electrons in the detector which is independent of the electron energy due to almost constant stopping power values for energies in the range of 0.2 and 20 MeV (see Fig. 4(a)). Therefore, the position of this bump is determined only by the thickness of the detector being transparent for high energy electrons. This effect can be observed in Fig. 4(b), where response functions to the IPC electrons/positrons are compared for different detector thicknesses. At very low energies of detected electrons (Figs. 3 and 5(b)) below 0.3 MeV, an additional spectrum structure can be observed which increases with the incident electron energy. As discussed in Section 3.3, this spectrum component results from electrons backscattered from the target and target holder. The stopping power and range of positrons in Si are about 2% lower than for electrons [1].

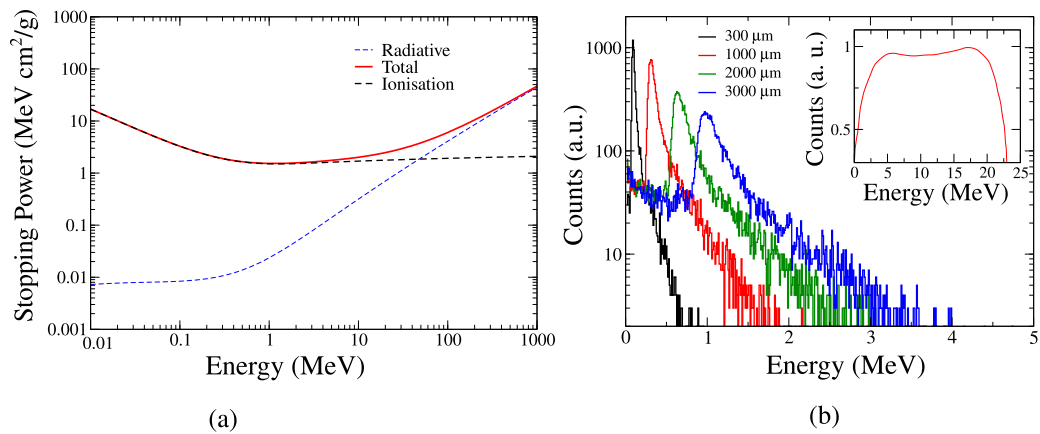

**Fig. 4.** (a) Energy dependence of the electron stopping power and its ionization and radiative components. (b) Response energy spectra of Si detectors with thicknesses of 0.3–3 mm obtained for electrons and positrons produced in the DD reaction, simulated with the Geant4 MC code. The original electron–positron energy spectrum employed in the simulation, originating from the theoretically predicted internal pair creation process [24], is presented in the inset.

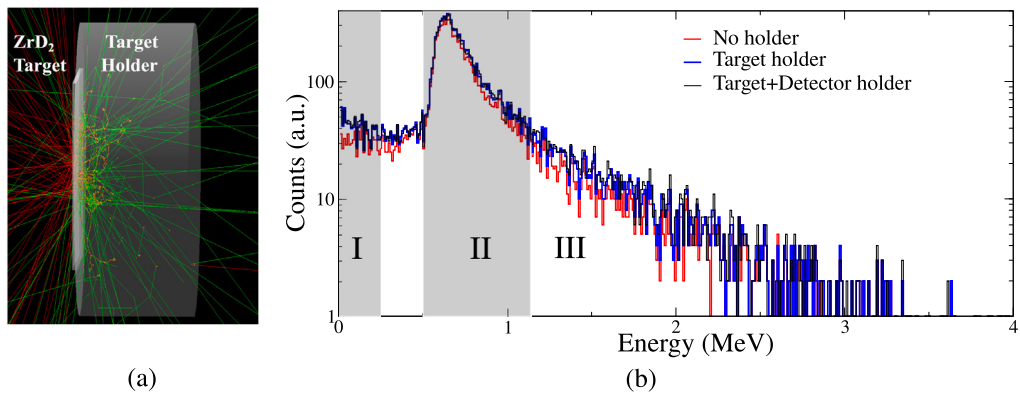

**Fig. 5.** (a) Electron scattering events within the target and target holder, simulated with Geant4 MC and visualized using OpenGL, electrons depicted in red and photons in green. (b) The response function of the 2 mm Si detector with a 46 μm Al absorber to the IPC electrons/positrons calculated using the Geant 4 code performed for the target without target holder (red), with target holder (blue) and additionally with the detector holder. The shaded areas represent the energy region of elastically scattered electrons (I) and the partial absorption peak (II), while the adjacent area beyond the region (II) is referred to as the high-energy tail (region III).

It means that the energy spectrum of detected positrons does not need to be simulated separately, and the number of emitted electrons can be simply increased by a factor of two.

*3.3. Effects of scattered electrons*

The backscattered electrons from the target holder and other parts of the experimental setup can have a significant impact on the precise measurement of the electron energy spectrum. Fig. 5(b) shows the Geant4 MC calculations of how backscattering from both the target holder and the detector holder affects the entire electron energy spectrum for a 2 mm detector with a 46 μm thick aluminium absorber. For a quantitative analysis, the energy spectrum is divided into three regions: the energy region of elastically scattered electrons and low-energy photons (I), the partial absorption peak region (II) and the high-energy tail region (III). The contribution from backscattering is separately calculated for all regions. The peak region (Region II in Fig. 5(b)) is of significant importance for experimentally determining the strength of the IPC transition. The high-energy tail region typically overlaps with other DD reaction products (protons, tritons, $^3$He) in the actual experiment (Fig. 7). Thus, additional absorption foils in the front of the detector can be applied to fully absorb or reduce the energy of heavy-charged particles while high-energy electrons will lose only a small fraction of their kinetic energy. In experiments, Al foils of thickness ranging between 0.8 – 125 μm were used. Unlike heavy-charged particles, the detection of electrons predominantly occurs within the entire volume of the detectors due to their much smaller stopping power values in silicon. The influence of detector dimensions (thickness and detection area) on the scattered electron contribution from the target-target holder system is thoroughly examined using simulations.

The multiple scattering in the thick target holder reduces the energy of the electrons and also produces bremsstrahlung photons. The low-energy electrons and photons collectively contribute to the increment in the lower energy region up to 200 keV across various detector geometries (Region I in Fig. 5(b)). The corresponding percentage increment of the counting rate in this energy range has been calculated as follows:

$$\% \ increment = \frac{A_H - A_0}{A_0} \times 100\%,$$

where $A_H$ is the area under the curve with a holder and $A_0$ is the area under the curve without holder. This is documented in Table 2 as the low energy increment.

The peak region is defined in the simulation as the range where the peak's height is reduced to 12% from its maximum, both to the left and right of the absorption peak. We have detailed this information in Table 2, where the increase in the peak area as a measure of backscattering was considered for detectors of various sizes. Notably, the present study reveals that backscattering from different sources within the experimental setup results in an average increase of about 18% in the overall spectrum for 0.3–3 mm thickness of the Si detector. Moreover, the analysis of the relationship between detector dimensions and the contribution of backscattering led us to the conclusion that detector dimensions do not significantly affect the detection of the backscattering components (as shown in Table 2). The position of the absorption peak becomes a crucial limiting factor, as depicted in Fig. 4(b). In thinner detectors e.g. ≤ 300 μm), the absorption peak blends with low-energy noise or other DD fusion products like $^3$H and $^3$He, making them indistinguishable.

Table 2
Results of Geant4 MC simulations of IPC electrons obtained for different Si detectors. An increase in counting rates in different regions of the energy spectrum due to scattered electrons is presented.

| Thickness (μm) | Surface area (mm$^2$) | Position of peak (MeV) | Peak to total area (%) | Low energy increment (%) | Peak area increment (%) | Tail area increment (%) | Total area increment (%) |
| --- | --- | --- | --- | --- | --- | --- | --- |
| 300 | 25 | 0.084 | 86 | — | 16 ± 3 | 18 ± 7 | 17 ± 3 |
| 1000 | 50 | 0.304 | 80 | 39 ± 9 | 17 ± 2 | 26 ± 5 | 19 ± 2 |
| 1000 | 100 | 0.305 | 82 | 30 ± 7 | 17 ± 2 | 29 ± 4 | 18 ± 1 |
| 2000 | 50 | 0.625 | 68 | 42 ± 9 | 14 ± 2 | 31 ± 5 | 19 ± 2 |
| 3000 | 50 | 0.980 | 62 | 55 ± 10 | 14 ± 5 | 26 ± 5 | 18 ± 2 |

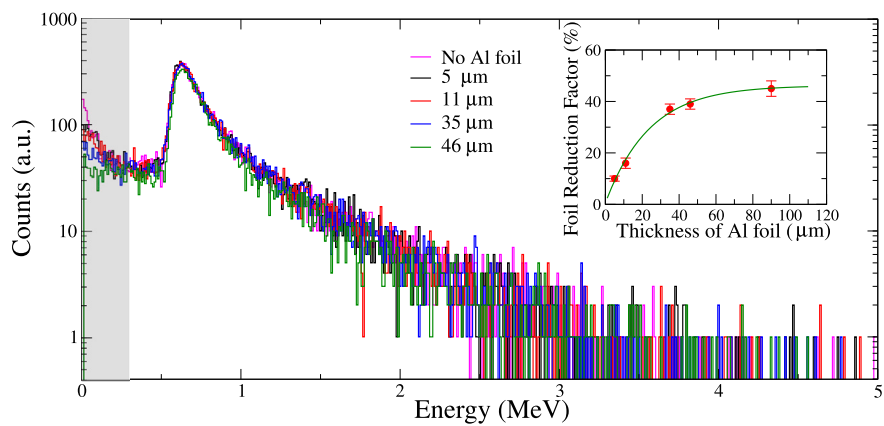

**Fig. 6.** Geant4 MC simulations of electron spectra performed for the 2 mm thick Si detector and for different thicknesses of Al absorption foil. The figure inset illustrates the corresponding percentage decrease in counts (Foil Reduction Factor) within the low energy region (up to 200 keV, marked in grey) for foils of varying thicknesses.

*3.4. Al absorption foil effect*

Aluminium absorption foils have a negligible impact on the high-energy electron (greater than 5 MeV) spectrum, but they play a critical role in counting electrons, protons, $^3$H, $^3$He particles in the entire energy spectrum of DD reactions, measured by a Si surface barrier detector. To accurately differentiate between different reaction channels of DD reaction in experiments at ultra-low energy, it is essential to keep other charged particle events separate from high-energy electron events. Fig. 6 illustrates electron energy spectra derived from the Geant4 MC simulation, using the 2 mm thick Si detector with different thicknesses of the Al absorption foils placed in the front of the detector. In the lower energy range (shaded region in Fig. 6), the variation in the number of counts is primarily attributed to differences in the thickness of aluminium foils. Thinner foils permit some low-energy, backscattered electrons to pass through, whereas thicker foils effectively block them. This effect, although not significantly affecting electron energy, does have a significant impact on the recorded counting rate. The position of the electron partial absorption peak remains, however, constant regardless of the foil thickness. For a more detailed analysis, a quantitative factor "Foils Reduction Factor" (FRF) is defined as the percentage decrease in counts with respect to the no-foil case observed in the low energy region (up to 200 keV, see Fig. 6) and presented in the inset of Fig. 6. The calculations performed for Al foil thicknesses ranging from 0.8 to 90 μm display a significant and conspicuous increase in FRF, reflecting a corresponding percentage decrease in counts. For foil thicknesses greater than 40 μm, the FRF increase becomes weaker.

In addition to the current Geant4 MC calculations, we also conducted calculations using the SRIM code [26] for various thicknesses of aluminium absorbers. These additional calculations have revealed that a 46 μm thick aluminium absorber foil offers the most favourable experimental conditions for observing and distinguishing the absorption peak of electrons and protons resulting from the DD reaction measured at ultra-low energies.

**4. Discussion and summary**

In the present paper, a measurement method of high-energy electrons using a single Si surface barrier detector has been presented. The thickness of the detector is much lower than the range of electrons so that, only a partial energy absorption spectrum can be observed. Since the stopping power value for electrons in the broad energy range 0.5 – 20.0 MeV is almost constant, we observe a broad partial energy absorption bump of the detected electrons, the position of which depends only on the thickness of the detector. Detailed Geant4 MC simulations, taking into account the entire experimental setup consisting of the target chamber, target holder, target plate and detector holder, show that the resulting energy spectrum for monoenergetic electrons is much more complicated. The study has been focused on the recognition of the energy spectrum component produced by the scattered electrons. This contribution is especially visible in the low energy part of the spectrum and arises from the backscattered electrons on the target holder. For the first time, we have shown that the strength of different energy spectrum contributions changes significantly for different detector thicknesses and absorption foils usually placed in the front of the detector to enable particle identification. The measurement method is especially useful for experiments dealing with high-energy electrons of a continuous energy spectrum, because of the absorption bump formation that allows to improve the effect-background ratio for processes of a relatively small probability.

An example of that is the internal $e^+e^-$ pair creation in the deuteron–deuteron (DD) nuclear reaction at energies far below the Coulomb barrier. This process, producing continuous energy electrons and positrons up to 22.8 MeV, is much weaker compared to the $^2$H(d,p)$^3$H and $^2$H(d,n)$^3$He reactions, even though an enhancement due to the recently observed threshold resonance in $^4$He can be expected [18,27]. To study this mechanism, the determination of the branching ratio between IPC and the proton emission for different deuteron energies is necessary [17]. To be sure that the observed energy bump in the detector energy spectrum (see Fig. 6) results from IPC, we have employed Si detectors of different thicknesses and used additional absorption Al foils in the front of the detectors. The measurements were accompanied by detailed Geant4 MC simulations which allowed us to determine the energy spectrum of emitted electrons /positrons and scattering effects studied before for monoenergetic electrons.

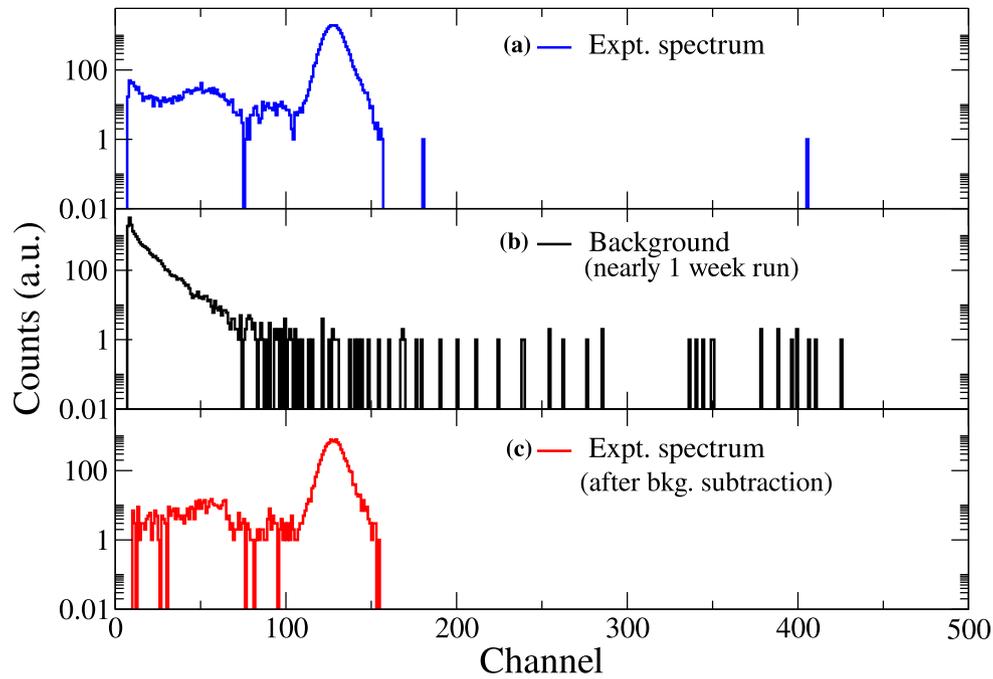

**Fig. 7.** (a) Experimental energy spectrum measured at deuteron energy 16 keV using the 2 mm thick Si detector and 46 μm thick Al absorption foil. (b) Experimental background for a long run. (c) Experimental spectrum after subtraction of the background.

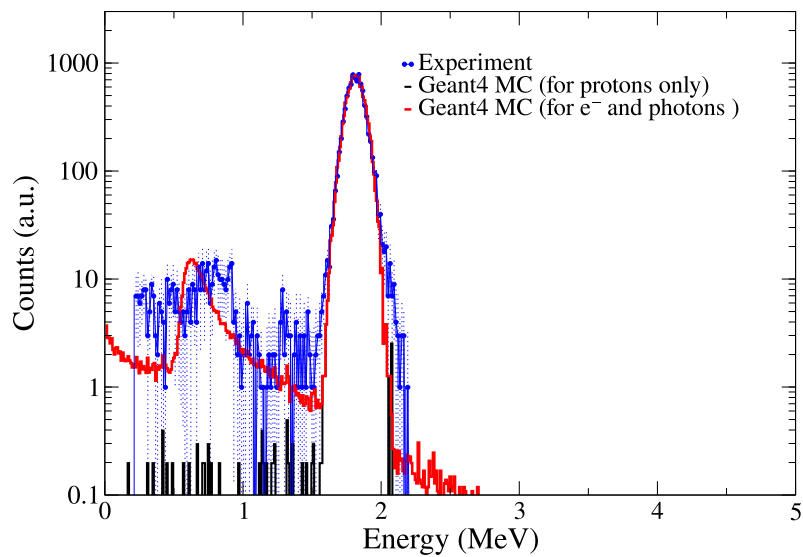

**Fig. 8.** Energy calibrated experimental spectrum after background subtraction (solid blue line with statistical error bars) obtained for 16 keV deuterons and the corresponding Geant4 MC simulation.

In Fig. 7, the charged particle spectrum measured by the 2 mm thick Si detectors with 46 μm-thick absorption Al foils at the deuteron energy of 16 keV is shown. The foil thickness was tuned to remove from the spectrum tritons of energy 1.01 MeV and $^3$He particle of energy 0.82 MeV. They are fully absorbed in the Al foil. The protons emitted with energy of about 3 MeV lose about 0.4 MeV in the Al foil and can be clearly seen with a maximum in channel 130. The measured long-run background (for almost a week) was subtracted from the originally measured spectrum, and a noticeable bump around channel 55, nearly ten times weaker than the proton peak could be observed. A corresponding Geant4 MC simulation and comparison with the experimental spectrum is presented in Fig. 8. The calculations have taken into account also the contribution resulting from the scattered protons which is of order $10^{-4}$ compared to the full energy line of protons. A slight difference in the calculated energy peak position of electrons (nearly 50 keV) can be observed. It might be due to uncertainty in straggling for high-energy electrons for Si medium and uncertainty of the actual thickness of Si surface barrier detector which depends on the supply voltage applied. An increase in the counting rate at very low energies corresponds mainly to the low-energy photons produced due to the bremsstrahlung and secondary recombination effects, e.g. about 80 keV X-rays induced in the gold surface layer of the detector.

**Table 3**
The electron–proton branching ratio (BR) determined, excluding all possible non-direct fusion events (electron scattered, proton scattered, Al absorber attenuation, low energy photons (LEP)) for the deuteron energies at 16 keV.

| BR @ 16 keV ($e^-/p$) | Reduction of BR (%) | | | |
|---|---|---|---|---|
| | Electron scattering | Proton scattering | Al absorber attenuation | Low energy photons (LEP) |
| $0.07 \pm 0.006$ | 0.05 | 0.01 | 0.001 | 0.003 |

The electron–proton ratio presented in Table 3 is an experimental count ratio measured using the 2 mm Si detector and determined according to the procedure described in Section 3.2. We integrated the count number in the energy region II (see Fig. 5(b), assigned only to electrons, corresponding to absorption peak in the energy spectrum. This number was divided by the number of protons measured at the same time to get the experimental value of the electron–proton ratio. The problem was, however, that the electrons are distributed over a much larger energy region, depending on the detector thickness. Using the Geant4 MC simulations, we could estimate how large was the fraction of the overall electron spectrum we took under consideration in region II and finally obtain the total value the electron–proton branching ratio.

Thanks to Geant4 MC simulations the branching ratio between the IPC and proton channel is about $0.07 \pm 0.006$, whereby all studied scattering effects have been included (see Table 3). The energy dependence of the branching ratio measured by different Si detector and Al foil thickness setups taking into account corresponding corrections could be also determined [17,28]. In summary, we would like to underline that the presented method enables to study very weak nuclear processes resulting in high-energy electron emission. However, an accompanying Monte Carlo simulation of different electron scattering processes in the actual applied experiment setup is necessary to determine the absolute strength of the studied phenomena.

### CRediT authorship contribution statement

**Gokul Das H:** Software, Resources, Formal analysis, Data curation. **R. Dubey:** Writing – review & editing, Writing – original draft, Visualization, Validation, Supervision, Software, Resources, Investigation, Formal analysis, Conceptualization. **K. Czerski:** Writing – review & editing, Validation, Resources. **M. Kaczmarski:** Resources. **A. Kowalska:** Resources. **N. Targosz–Ślęczka:** Resources. **M. Valat:** Resources.

### Declaration of competing interest

The authors declare that they have no known competing financial interests or personal relationships that could have appeared to influence the work reported in this paper.

### Data availability

Data will be made available on request.

### Acknowledgements

This project has received funding from the European Union's Horizon 2020 research and innovation programme under grant agreement No 951974.